\documentclass[12pt,fleqn]{article}
\usepackage{amsmath,a4,latexsym,epsfig}
\newcommand{\intd}{\int \! d^4 x \;}
\newcommand{\intS}{\int \! d S \;}
\newcommand{\intSbar}{\int \! d \bar S \;}
\newcommand{\intV}{\int \! d V \;}
\newcommand{\Ga} {\Gamma}
\newcommand{\Gacl} {{\Gamma_{\rm cl}}}

\renewcommand{\L}{{\cal L}}
\newcommand{\Abar}{{\bar A}}
\newcommand{\lambdabar}{{\overline\lambda}}
\newcommand{\sigmabar}{{\overline\sigma}}
\newcommand{\psibar}{{\overline\psi}}
\newcommand{\Fbar}{{\bar F}}
\newcommand{\phibar}{{\overline \varphi}}
\newcommand{\epsilonbar}{{\overline\epsilon}}
\newcommand{\thetabar}{{\overline\theta}}
\newcommand{\etabar}{{\overline\eta}}
\newcommand{\chibar}{{\overline\chi}}
\newcommand{\qbar}{{\overline q}}
\newcommand{\fbar}{{\overline f}}
\newcommand{\cbar}{{\overline c}}

\newcommand{\alphadot}{{\dot\alpha}}

\newcommand{\lambdaV}{{\tilde \lambda}}
\newcommand{\lambdaVbar}{{\overline {\tilde \lambda}}}
\newcommand{\DV}{{\tilde D}}
\newcommand{\Tr}{{\rm Tr}}

\def\dF#1{\frac{\delta{\cal F}}{\delta#1}}

\def\df#1{\frac{\delta}{\delta#1}}

\def\pslash#1{{\setbox0=\hbox{$#1$}
  \rlap{\ifdim\wd0>.7em\kern.22\wd0\else\kern.1\wd0\fi /}#1}}

\def\brs{\mathbf s}
\newcommand{\mn}{\mu\nu}

\begin{document}
\begin{titlepage}

\begin{flushright}
BN--TH--2002--01\\
DESY--02--010\\
{\tt hep-ph/0201247}\\
\end{flushright}
\vspace{8ex}
\begin{center}
{\large\bf{
        Softly broken supersymmetric Yang-Mills theories:\\[1ex]
         Renormalization and non-renormalization theorems}}
\\
\vspace{8ex}
{\large       E. Kraus$^a$} and
{\large        D. St{\"o}ckinger$^b$}
{\renewcommand{\thefootnote}{\fnsymbol{footnote}}
\footnote{E-mail addresses:\\
                kraus@th.physik.uni-bonn.de,\\
                dominik@mail.desy.de.}} 
\\
\vspace{2ex}
{\small\em               $^a$ Physikalisches Institut,
              Universit{\"a}t Bonn,\\
              Nu{\ss}allee 12, D--53115 Bonn, Germany\\}
\vspace{.5ex}
{\small\em $^b$ Deutsches Elektronen-Synchrotron DESY,\\
                D--22603 Hamburg, Germany\\ }
\vspace{2ex}
\end{center}
\vfill
{\small
 {\bf Abstract}
 \newline\newline
We present a minimal version for the renormalization of  softly broken
Super-Yang-Mills theories using the extended model with a local
gauge coupling. It is shown that the non-renormalization theorems of
the case with unbroken supersymmetry are valid without modifications
  and that the
renormalization of the soft-breaking parameters is completely governed
by the renormalization of the supersymmetric parameters. 
 The 
symmetry identities in the present context are peculiar, since
the extended model contains two anomalies:  the Adler-Bardeen anomaly
of the axial current and  an anomaly of supersymmetry in the presence
of the  local gauge coupling. From the anomalous symmetries we derive
the exact all-order expressions for the $\beta$ functions of the gauge
coupling and of the  soft-breaking parameters. They generalize earlier
results to arbitrary normalization conditions and imply the NSVZ
expressions for a specific normalization condition on the coupling.

\vfill
}

\end{titlepage}


\newpage
\section{Introduction}

Renormalization of softly broken supersymmetric gauge theories is an
important issue for  a better understanding of supersymmetry and its
possible realization in nature. Proofs of
the all-order renormalizability of these  theories in
the Wess-Zumino gauge have
been proposed in refs.~\cite{MPW96b} and \cite{HKS00}.
%
However, renormalization of softly
broken supersymmetric theories remains unsatisfactory in both
approaches. There are far-reaching relations and non-re\-nor\-ma\-lization
theorems that govern the renormalization of the supersymmetric and the
 soft-breaking parameters, but these relations are not
prescribed by the symmetries used in \cite{MPW96b,HKS00} and thus
escape the algebraic renormalization procedure. 

The existence of    relations between the renormalization of softly broken and
supersymmetric theories has been suggested by \cite{YA94} and has
been further elaborated in refs.~\cite{AKK98, KAVE00,KO00} 
 using superfield techniques
in superspace. 
Most of the corresponding calculations have been performed assuming
a supersymmetric and gauge-invariant regularization scheme. 
But that these symmetries    by themselves do not lead to the
special renormalization properties
is seen from the algebraic
renormalization of soft supersymmetry-breakings in the spurion
formalism
\cite{HKS00}, where the symmetries are accounted for but restrictions
on parameters    are not present. Therefore, in these old approaches
it was difficult to identify those aspects of the symmetries that are
the underlying reasons for the special renormalization properties. 

The renormalization properties    of the soft-breaking parameters are
close to the non-renormalization theorems in models with unbroken
supersymmetry
\cite{FULA75,GSR79}. These  non-renormalization theorems 
have been first derived  in superspace, but not as a direct   
consequence of supersymmetry. Recently it has been shown in a series of
papers \cite{FLKR00,KRST01,KRST01soft,KR01}
that the non-renormalization theorems are in fact consequences of the
special structure of supersymmetric Lagrangians. Every supersymmetric
part of the
Lagrangian is the highest component of a supermultiplet and thus
can be expressed as the supersymmetry variation of a
lower-dimensional field polynomial.    
This aspect is the origin of the improved
renormalization properties in supersymmetric theories.

The way to make use of the multiplet 
structure of supersymmetric Lagrangians is to couple
the multiplets to
external fields. These external fields are chiral or vector
supermultiplets according to the structure of the respective
Lagrangian multiplets, and 
they   
have an immediate
interpretation (see also
\cite{AGLR98}): The external field coupled to the Super-Yang-Mills
part corresponds to an extension of the coupling to an external
superfield and the external multiplet coupled to the matter part is the
axial vector multiplet,  whose vector component couples to the axial
$U(1)$ current. 
Hence, the derivation of non-renormalization theorems is traced back to
renormalization of supersymmetric field theories in presence of local
couplings and by taking into account axial symmetry. 

It is obvious that this construction intrinsically includes soft
breakings of the Girardello-Grisaru class \cite{GIGR82}, since these
breakings are  
just the lowest components of the Lagrangian multiplets and are
automatically    present
in the extended construction of supersymmetric theories with local
coupling.  The purpose of the present paper is to work out
the consequences of the extended construction for softly broken
supersymmetry. It will indeed turn out that  
the divergency structure is unchanged by passing from
the unbroken to the softly broken case, and we will find explicit
results for the specific renormalization properties of softly broken
theories.

Renormalization of supersymmetric Yang-Mills theories in presence of the 
local coupling has to be performed with care: We have shown that not
only the axial current is anomalous \cite{AD69}, but that also supersymmetry has
an anomaly \cite{KR01,KR01anom}. Both anomalies can be consistently absorbed into redefined 
symmetry identities, the Adler-Bardeen anomaly by an anomalous axial
transformation  of  the space-time dependent $\Theta$ angle of QCD,
 the supersymmetry anomaly by an
anomalous supersymmetry transformation of the local gauge coupling.
The resulting    symmetry identities govern the
renormalization of Super-Yang-Mills theories with soft breakings to
all orders and imply the non-renormalization theorems as well
as the relations between soft-breaking  and supersymmetric
parameters.

The most important implications of the    anomalous symmetries are
their effects on the renormalization group functions.
The supersymmetry anomaly induces the two-loop term in the
$\beta$ function of pure Super-Yang-Mills theories, whereas the
Adler-Bardeen anomaly induces the matter part contributions to the
gauge-$\beta$ function. Finally, the complete all-order construction 
implies    a closed
expression for the gauge-$\beta$ function \cite{KRST01,
KR01}, which takes with particular normalization  conditions the NSVZ form
\cite{NSVZ83}.
Similarly, we prove here that the anomaly coefficients show up in the
$\beta$ functions of the gaugino mass \cite{HISH98,JAJO97gaugino}
and    
in the $\beta$ function of the scalar mass
\cite{JAJO98scalar,KKZ98,JAJO98scalar2}. 
These results also clarify the reason of
the well-known    failure of superfield techniques for the derivation of the
scalar-mass $\beta$ function.

The analysis of the present paper is similar to the one we have
carried out in SQED \cite{KRST01soft}, but it differs in details due to the
appearance of the  supersymmetry anomaly \cite{KR01}. Hence, we give
all results in a very condensed form and refer for technical details
to \cite{KRST01soft}. 
In section 2 we present the classical supersymmetric and gauge
invariant action with local coupling and soft supersymmetry-breaking.
As a direct consequence, the non-renormalization theorems and the
relations for the soft-breaking parameters are formulated on the
level of the invariant counterterms in section 3.
In section 4 we renormalize the model  by
constructing
the  anomalous
Slavnov-Taylor identity that governs the renormalization of softly
broken Super-Yang-Mills theories in the Wess-Zumino gauge.
 In section 5 we derive the closed expressions of soft
$\beta$ functions from the
symmetry identities.   Generalizations to more complicated models and
discussions of the results can be found in the conclusions.
In the appendix  we summarize the BRS transformations of the fields in
presence of local couplings. Notations  and conventions are the ones
given in appendix A of \cite{KRST01}.

\section{The extended model}

The basic observation behind the non-renormalization theorems is that
every supersymmetric part in a supersymmetric Lagrangian is the
highest component of 
a supermultiplet. This fact can be made visible in an extended model
of Super-Yang-Mills (SYM) theories \cite{KR01}, where three kinds of
external supermultiplets are introduced that couple to 
the different parts of the Lagrangian. The most important of these is
the local gauge coupling $g(x)$ with its supermultiplet $G(x, \theta, \thetabar)$:
\begin{eqnarray}
\label{E2def}
G (x, \theta, \thetabar ) & = & ({\mbox{\boldmath{$\eta$}}}(x, \theta,
\thetabar ) +{\mbox{\boldmath{$\etabar$}}}(x, \theta, \thetabar ))^ 
{-\frac 12} \equiv g(x) + {\cal O}(\theta, \thetabar)\ .
\end{eqnarray} 
Due to the chiral structure of the SYM action the
multiplet of the local coupling is decomposed into 
a dimensionless  chiral and antichiral field multiplet,
 ${\mbox{\boldmath{$\eta$}}}$ and 
${\mbox{\boldmath{$\etabar$}}}$, with the following component expansion in chiral and antichiral representation, respectively:
\begin{equation}
\label{defeta}
{\mbox{\boldmath{$\eta$}}}(x, \theta) = \eta + \theta^ \alpha
 \chi_\alpha + \theta^2 f \ ,
 \qquad 
{{\mbox{\boldmath{$\etabar$}}} }(x, \thetabar) = \etabar + \theta_\alphadot \chibar ^
 \alphadot  + \thetabar^2 {\overline f} \ .
\end{equation}
As   shown in refs.~\cite{KRST01,KR01} 
the extension to local coupling requires at the same time to introduce
the multiplet of the
axial vector field,
\begin{equation} 
\label{Vdef}
\phi^V =   \theta \sigma ^ \mu \thetabar V^ {\mu} -i \thetabar^2 \theta^ \alpha
 \lambdaV_ \alpha + i \theta^2 \thetabar_\alphadot \lambdaVbar^
 \alphadot + \frac 14  \theta^2 \thetabar ^2 \DV\  ,
\end{equation}
whose vector component $V^\mu$ couples to the anomalous axial $U(1)$ current,
and a chiral and an antichiral multiplet, 
$\mathbf q$ and $ \mathbf \qbar$, which couple to the matter mass term
and govern the soft breaking of axial symmetry:
\begin{equation}
\label{qdef}
{\mathbf q} =
q + \theta q^ \alpha + \theta^2 q_F \quad \mbox{and} \quad
{\mathbf \qbar} =
 \qbar + \thetabar _\alphadot\qbar^ \alphadot + \thetabar^ 2 \qbar_F.
\end{equation} 

The classical action
of SYM with local coupling  is decomposed into the supersymmetric
Yang-Mills part, the
matter part and the matter mass term and $q$-field interaction,
\begin{equation}
\label{Gasusy}
{\Ga_{\rm susy}} = \Ga_{\rm YM} + \Ga_{\rm matter} + (\Ga_{\rm mass} +
\Ga_q)\ . 
\end{equation}
The individual contributions are invariant under non-abelian gauge
 transformations, supersymmetry and softly broken $U(1)$ axial
 symmetry. For simplification
we restrict the construction to non-abelian gauge theories with a
Susy-QCD-like structure. We assume a simple gauge group with
 generators $\tau_a$ and consider
 left- and right-handed matter fields, each in an irreducible
 representation generated by $T_a$ and $(-T_a{}^*)$, respectively.  

The contributions to $\Gamma_{\rm susy}$ can be efficiently written by
   using the following superfield expressions for the gauge multiplet $\phi^A$
and for 
the left- and right-handed matter fields $A_L,A_R$ and  $\Abar_L, \Abar_R$:
\begin{eqnarray} 
\label{Adef}
\phi^A &= &  \theta \sigma ^ \mu \thetabar A^ {\mu} -i \thetabar^2 \theta^ \alpha
 \lambda_ \alpha + i \theta^2 \thetabar_\alphadot \lambdabar^
 \alphadot + \frac 14  \theta^2 \thetabar ^2 D \ , \qquad \phi^A =
 \phi^A_a \tau_a\ ,\\
A_X & = & \varphi_X + \psi_X \theta + F_X  \theta^2\ ,
 \quad 
\Abar_X   = 
 \phibar_X  + \psibar_X  \thetabar +
 \Fbar_X  \thetabar ^2\ ,\quad 
X=L,R,
\end{eqnarray}
with the normalization $\Tr (\tau_a\tau_b)=\delta_{ab}$ for the
generators.   
Expressed in terms of superspace integrals, the invariant classical action in
the Wess-Zumino gauge reads   
\begin{eqnarray}
\label{GaYM}
\Ga_{\rm YM} &= &- \frac 14\intS {\mbox{\boldmath{$\eta$}}}  
{\cal L}_{\rm YM}
- \frac 14 \intSbar {{\mbox{\boldmath{$\etabar$}}}}  \bar {\cal L}_{\rm YM} \ ,\\
\label{Gamatter}
\Ga_{\rm matter} & = & \frac 1 {16}\intV \L_{\rm matter}\ , \\
\label{Gamass}
\Ga_{\rm mass} + \Ga_q &= & 
- \frac 1 4\intS ({\mathbf q} + m ) \L_{\rm mass}
- \frac 1 4\intSbar ({\mathbf \qbar} + m ) \bar \L_{\rm mass}\ .
\end{eqnarray}
with  
\begin{eqnarray}
\label{LYM}
\L_{\rm YM} &= & \Tr W^ \alpha W_{\alpha } = W^ \alpha_a W_{\alpha a } \ ,  \quad
W_\alpha \equiv  \frac 1 {8\sqrt 2 } \bar D \bar D (e^ {-2g(x) \phi^ A } D_\alpha 
e^ {2g(x) \phi^ A })\, ; \\  
\label{Lmatter}
\L_{\rm matter} & = &  \Abar_{Lk} \left(e^ { 2 g \phi^ A_a T_a + 2
\phi^ V }\right)^k{}_j A^ j_L +
A_{Rk} \left(e^ {- 2 g \phi^ A_a T_a + 2 \phi^ V }\right)^k{}_j \Abar^ j_R  \ ,\\
\label{Lmass}
\L_{\rm mass}  &= & 
 A_R^k A_{L k}\ .
\end{eqnarray}
Gauge and supersymmetry transformations of the individual fields
can be read off from the
corresponding BRS transformations of the appendix  (see (\ref{BRS})).
In  the limit to constant supercoupling $G(x, \theta, \thetabar) \to
g=$ const.\ we obtain 
  the usual classical action of SYM theories  with gauged axial symmetry.

The SYM action with local coupling can immediately be extended to a consistent
description of softly broken supersymmetry. The external fields have  
exactly the properties of the spurion fields used in \cite{GIGR82} to
describe soft breaking. Indeed, the only change  we have to apply
to the model of \cite{KR01} is to shift
 the highest components of the external multiplets
by constant mass parameters,
\begin{eqnarray}
\label{massshifts}
& & f \to f + \frac {M_\lambda} {g^2} \ ,
\qquad
\bar f \to  \bar f + \frac {M_\lambda}{ g^ 2} \ ,
\nonumber\\
& & q_F \to q_F - b \ , \qquad \qbar_F \to \qbar_F - b , \nonumber \\ 
& & \tilde D  \to   \tilde D - 2 M^2   .
\end{eqnarray}
Through these shifts all parity conserving soft-breaking terms of the Girardello-Grisaru
class \cite{GIGR82} are generated in the classical action,    i.e.
\begin{equation}
\Ga_{\rm susy} \to \Ga_{\rm susy} + \Ga_{\rm soft}
\end{equation}
with
\begin{eqnarray}
\label{Gasoft}
\Ga_{\rm soft} &= &\intd \Bigl(-\frac 12 M_\lambda (\lambda \lambda +
\lambdabar \lambdabar) \nonumber \\
& & \phantom{\intd} - M^ 2 (\varphi_L \phibar_L + \varphi_R \phibar_R)
- b( \varphi_L \varphi_R + \phibar_L \phibar_R)\Bigr) \ .
\end{eqnarray}
If the shifts (\ref{massshifts}) are
also included in the supersymmetry transformations and
in the axial transformations of the external fields,
 the softly broken model is characterized
by the same symmetries as the original unbroken model.

The crucial point of the present construction of soft
 supersymmetry-breaking are the highly restrictive symmetry properties
 of the 
 extended model.   They give access not only to the
 non-renormalization theorems \cite{KR01} but also to the
 renormalization properties of the soft-breaking parameters.  

\section{Improved renormalization properties}
\label{sec:RenProperties}

The explicit form of the divergences in any quantum field theory generally
depends on the regularization scheme.
However, after the possible
symmetry-breaking effects of the regularization are cancelled by suitable
counterterms, the structure of the remaining divergences is
scheme-independent. These divergences are the superficial divergences
and
correspond order by order to local field
monomials that are invariant with respect to the classical
symmetries. Hence, they can be cancelled by invariant
counterterms and an investigation of invariant counterterms
is equivalent to an investigation of the structure of the divergences.

The properties of the extended model put severe constraints on the
invariant counterterms of physical parameters. 
These  terms are characterized by being invariant under the same
symmetry transformations
 as $\Gamma_{\rm susy}$
 and can be derived already
at the present stage.
 The results will exhibit all the special
renormalization properties of softly broken SYM theories.
 We will continue the discussion of invariant
counterterms  in sec.\ \ref{sec:invct}. There
we will find  further  invariant counterterms, which
correspond to field renormalizations and include a redefinition of the
classical symmetry transformations.

In order to obtain full control over the renormalization, it is
necessary to use two further constraints.  On the one hand, $(\eta
+\etabar)^{-1/2}$ is identified with the local gauge coupling
(\ref{E2def}). 
Since for constant gauge coupling the loop expansion is a power series in
the coupling and an $l$-loop counterterm is of the order $g^{2l}$, this
identification fixes the powers of ${\mbox{\boldmath{$\eta$}}}$
 in the counterterms (see
(27) later for an explicit formula).
 On the
other hand,  
 $(\eta - 
\etabar)$ takes the role of a space-time dependent $\Theta$ angle of
the model.
Thus, the classical action depends on
 it only via a total derivative, and this property can be expressed by 
 the following identity:
\begin{equation}
\label{holomorphcl}
\intd \Bigl(\df{\eta} - \df{\etabar}\Bigr) \Ga_{\rm susy} = 0\ .
\end{equation}
The same identity has to hold for the invariant counterterms.

Using furthermore an additional R-invariance of the classical action (for the
explicit form see (\ref{WIR}), (\ref{WIRoperator}))
the results for the gauge invariant and supersymmetric field
monomials of 
$l$-loop order are
given by
\begin{eqnarray}
\label{Gactphysgen}
\Ga^{(l)}_{\rm ct,phys} & = &
- \hat z_{\rm YM}^ {(l)} \frac 14\intS  \Big({\mbox{\boldmath{$\eta$}}}+
\theta^2 \frac {M_\lambda} {g^2}\Big)^ {-l+1}  
{\cal L}_{\rm YM} + \mbox{c.c.}\nonumber \\ 
& & - \hat z_{\rm mass}^ {(l)}\frac 14 \intS \Big({{\mbox{\boldmath{$\eta$}}}}+ \theta^2 \frac
{M_\lambda} {g^2}\Big) ^{-l}
({\mathbf q} + m - \theta^2 b) \L_{\rm mass} + \mbox{c.c.} \nonumber \\
& & +\frac 1 {16} \intV
{\cal K}^ {(l)} \Bigl({\mbox{\boldmath{$\eta$}}} +
\theta^2 \frac {M_\lambda} {g^2},
{\mbox{\boldmath{$\etabar$}}}
+\thetabar^2 \frac {M_\lambda} {g^2}\Bigr)   \big(1- \theta^2 \thetabar^2 M^2
\big) \L _{\rm matter}  ,
\end{eqnarray}
with the restriction ${\cal K}^
{(l)}({\mbox{\boldmath{$\eta$}}},{\mbox{\boldmath{$\etabar$}}}) \to 
\hat z_{\rm matter} ^ {(l)} g^{2l}$.

Applying the identity (\ref{holomorphcl}) on $\Gamma_{\rm ct,phys}$ 
we find the following constraints that have to be satisfied by the invariant
counterterms:
\begin{eqnarray}
\label{zrestricted}
& & \hat z_{\rm YM}^ {(l)} = 0 \quad \mbox{for}\quad l\geq 2 \ ,\qquad
 \hat z_{\rm mass}^ {(l)} = 0 \quad \mbox{for} \quad l\geq 1 \
 ,\nonumber \\
& & {\cal K}^ {(l)} ({\mbox{\boldmath{$\eta$}}},
{\mbox{\boldmath{$\etabar$}}})
 =  \hat z_{\rm matter} ^ {(l)}
({\mbox{\boldmath{$\eta$}}} + {\mbox{\boldmath{$\etabar$}}})^ {-l}\ .
\end{eqnarray}

Hence,   
the  counterterm to the Yang-Mills part is restricted to one-loop
order and
 invariant counterterms to
the supersymmetric mass term are absent. These restrictions on
physical counterterms state the 
non-renormalization theorems of chiral vertices \cite{FULA75,GSR79}
and the generalized
non-renormalization theorem of the coupling constant \cite{SHVA86}.
In addition,
the renormalization constants of softly broken
parameters  are related to the renormalization
constants of supersymmetric parameters. 

For constant coupling it is possible to express the invariant
counterterms as field and parameter renormalizations. For the
 $z$-factors of the supersymmetric parameters, the coupling and the
fermion mass parameter,  one obtains in loop order $l$: 
\begin{eqnarray}
z^ {(1)}_{g^2} &  = &  -   g^ 2 z^ {(1)}_{\rm YM} \quad \mbox{and} 
  \quad z^ {(l)}_{g^2} = 0  \quad \mbox{for}\quad  l \geq 2\ ,\nonumber \\
z^ {(l)}_m  &  =  & - 2 g^ {2l} z_{\rm matter}^ {(l)} \ .
\label{zsusy}
\end{eqnarray}
The $z$-factors of the soft mass parameters are  determined as
functions of $z_{g^2}$ and $z_m$:
\begin{eqnarray}
\label{zsoft}
%
z^ {(1)}_{M_\lambda} &= &     z^ {(1)}_{g^2}  \quad \mbox{and} \quad
z^ {(l)}_{M_\lambda} =    0 \quad \hbox{for} \quad l\geq 2,
\nonumber \\
z^ {(l)}_b & = &  \Big(2l \frac {M_\lambda m} b +1\Big) z_m^ {(l)} \ ,\quad
z^ {(l)}_M =  \frac 12  l(l+1) \frac {M_\lambda^ 2} {M^ 2} z_m^ {(l)}
\ .
\end{eqnarray}  
The renormalization constants
to the soft-breaking scalar masses match the relations derived first by
\cite{YA94} and later on  by \cite{AKK98} in superspace.    The
renormalization constants to the gaugino  mass are related to the ones
of the gauge coupling and are further restricted
by the non-renormalization theorem of the gauge coupling in a similar
way as proposed in ref.~\cite{HISH98}, where  
 holomorphicity was imposed as
an additional constraint on chiral
 invariant counterterms. In the present  
approach the constraint is a consequence of the identity~(\ref{holomorphcl}).

The results for the counterterms reflect the improved
renormalization-behaviour and the special divergency structure of SYM
theories. However, unlike in many other models, $\beta$ functions
cannot be inferred  from the $z$-factors (\ref{zsoft}) and
(\ref{zsusy}). As we will see later, in the course of renormalization
the symmetries are broken by two anomalies: Axial symmetry is broken
by the Adler-Bardeen anomaly \cite{AD69} and
supersymmetry by a supersymmetry anomaly \cite{KR01}.
 For this reason an invariant regularization scheme for the extended
model cannot exist, and a discussion of the $\beta$ functions based on
invariant counterterms cannot be performed. 
A derivation of the $\beta$ functions requires an algebraic
construction based on the anomalous symmetries of the model.

 \section{Quantization}

\subsection{The classical action and its symmetries}

For quantizing supersymmetric gauge theories in
the Wess-Zumino gauge the symmetry transformations
 of the model, gauge transformations,
$U(1)$ axial transformations, supersymmetry and translations are 
summed up in the BRS trans\-for\-mations
\cite{White92a,MPW96a,HKS99,KRST01,KR01}:
\begin{equation}
\label{BRS}
\brs \phi = (\delta^ {\rm gauge}_{c(x)} + \delta^ {\rm axial}_{\tilde
c(x)} + \epsilon ^ \alpha \delta_\alpha + \bar \delta_\alphadot
\epsilonbar^ \alphadot - i \omega^ \mu \partial_\mu) \phi \ .
\end{equation}
The ghost fields $c(x), \tilde c(x)$  replace the local
transformation parameters of gauge transformations and axial
transformations, and the constant ghosts $\epsilon^ \alpha $,
$\epsilonbar^ \alphadot$ and $\omega^ \mu$ are the constant supersymmetry
 and translational ghosts, respectively.
BRS transformations of the ghosts are determined by the structure
constants of the algebra and the algebra of symmetry transformations
is expressed in the on-shell nilpotency of the BRS operator. 
 The BRS transformations of the
fields (with auxiliary fields being eliminated) are collected    in
appendix A.
 They differ from the BRS
transformations in the symmetric model by  shifts
in the scalar components of external fields according to (\ref{massshifts}).

By means of BRS transformations it is possible to add a BRS invariant
gauge-fixing and ghost part to the action  
\begin{equation}
\label{Gagf}
\Ga_{\rm g.f.} = \brs \Tr\intd \Bigl( \frac 12
\xi \bar c B + \bar c  {\cal F} \Bigr)
=\Tr\intd \Bigl( \frac 12
\xi   B^ 2 + B {\cal F}\Bigr) + \Ga_{\rm ghost} \ .
\end{equation}
The fields $B$ are the Lagrange multiplier fields, and the function
${\cal F}$ describes  an appropriate linear gauge fixing function
for the longitudinal part of the gauge fields:  ${\cal F} = \partial A
+ \dots $.

Finally, the BRS transformations $\brs\phi$ that are non-linear in the
propagating fields are coupled to external fields $Y_\phi$, and the
external field part $\Gamma_{\rm ext.f.}=\int Y_\phi \brs\phi$ is added to the
classical action, so that  
\begin{equation}
\label{Gacl}
\Gacl = \Ga_{\rm {susy}} + \Ga_{\rm soft}
+ \Ga_{\rm g.f.} + \Ga_{\rm ext. f.} \ .
\end{equation}
From this complete classical action the auxiliary fields are
eliminated using their equations of motion.  
In this procedure, bilinear expressions in external fields are induced
that compensate  the equations-of-motion terms in the supersymmetry algebra.

The BRS invariance of $\Gacl$ can be rewritten in the form of
the Slavnov-Taylor identity
\begin{equation}
\label{ST}
{\cal S} (\Gacl)  = 0\ ,
\end{equation}
with the usual
bilinear Slavnov-Taylor operator (see eq.\ (\ref{SToperator})). 
  

The dependence of $\Gacl$ on the field multiplets $
{\mbox{\boldmath{$\eta$}}}$ and ${\mbox{\boldmath{$\etabar$}}}$ is 
constrained by the identity  
(\ref{holomorphcl})
\begin{equation}
\label{holomorphcl2}
\intd \Bigl(\df{\eta} - \df{\etabar}\Bigr) \Gacl = 0\ ,
\end{equation}
 and by  identifying the real part
of the lowest components $\eta,\etabar$
 with the local gauge coupling (\ref{E2def}). Direct inspection of the diagrams
shows that this identification leads to the  following 
 topological formula for one-particle irreducible (1PI)
 diagrams in loop order $l$:
\begin{equation}
\label{topfor}
N_{g(x)} = N_{\rm amp. legs} + N_Y+ 2N_f+
2N_\chi + 2 N_{\eta - \etabar} 
+ 2(l-1)\ ,
\end{equation}
Here  $N_{\rm amp. legs} $ counts the number of external
amputated legs with propagating fields ($A^ \mu, \lambda, \varphi_A,
\psi_A, c, \cbar  $ and the respective
complex conjugate fields), $N_Y$ gives the number of BRS insertions,
counted by the number  of differentiations with respect to the
external fields $Y_\phi$. 
 $N_f$, $N_\chi$ and $N_{\eta - \etabar } $ give the number of 
insertions 
corresponding to  the respective external fields. 
The validity of the topological formula in higher orders  
ensures that the limit to constant coupling results in the 1PI Green
functions of ordinary SYM theories with soft breaking. 

Furthermore, the classical action is   restricted by a  softly broken
R-symmetry, which is defined in the same way as in SQED
\cite{KRST01soft}. We impose the corresponding Ward identity not only
for $\Gacl$ but also for the full vertex functional $\Gamma$:  
\begin{equation}
\label{WIR}
{\cal W}^ R \Ga  = 0
\end{equation}
with
\begin{eqnarray}
\label{WIRoperator}
{\cal W}^ R  & = &  i \intd \biggl( \sum_{A= L,R}\Bigl(\varphi_A
\frac {\delta} {\delta \varphi_A }  -
Y_{\varphi_A}
\frac {\delta} {\delta Y_{\varphi_A} }\Bigr) +
  {\lambda^ \alpha}
\frac{\delta}{\delta \lambda^\alpha} -
Y_\lambda^ \alpha
\frac {\delta} {\delta Y_\lambda^ \alpha } \nonumber \\ 
 & & \phantom{i \intd} { }  +
{\lambdaV ^ \alpha}\frac{\delta}{\delta \lambdaV^\alpha}
-  q ^\alpha \frac {\delta} {\delta q ^ \alpha} - \chi^ \alpha
\frac{\delta}{\delta \chi^ 
\alpha} \nonumber \\
& & \phantom{i \intd} { } 
 - 2 (q_F -b)  
\frac {\delta}{\delta q_F} 
 - 2 \Big(f + \frac {M_\lambda} {g^2}\Big)
\frac{\delta}{\delta f} - {\rm c.c.} \biggr)  \nonumber  \\
& &{ } +  i \Bigl( \epsilon^ \alpha \frac{\partial}{\partial
\epsilon^ \alpha} - \epsilonbar^ \alphadot \frac{\partial}{\partial
\epsilonbar^ \alphadot} \Bigr) \ . 
\end{eqnarray} 
It implies invariance under R-parity, where all superpartner-fields
are transformed to their negative.

\subsection{Renormalization and anomalies}

In one-loop order,  axial symmetry is broken by the
Adler-Bardeen anomaly \cite{AD69,ADBA69}. And when we impose the identity
(\ref{holomorphcl2}) to all orders\footnote{
 The supersymmetry anomaly could be shifted
to the identity (\ref{holomorph}) but implies there the unwanted
feature of the renormalization of the $\Theta $ angle of QCD ($ 2 \eta \equiv
\frac 1 {g^2} + i \Theta$) (see \cite{KR01}).}, i.e.,
\begin{equation}
\label{holomorph}
\intd \Bigl(\df{\eta} - \df{\etabar}\Bigr) \Ga = 0\ ,
\end{equation}
 then supersymmetry is broken by a
supersymmetry 
anomaly \cite{KR01}. Therefore the Slavnov-Taylor identity is anomalous in
one-loop order:
\begin{equation}
\label{STanomalous1loop}
{\cal S} (\Ga) = r^{(1)} \Delta^{\rm axial} + r_{\eta}^ {(1)} 
\Delta^{\rm susy} + {\cal  O}(\hbar^2)\ .
\end{equation}
The anomalous field monomials are given by the following expressions:
\begin{eqnarray}
\label{ABanomaly}
\Delta^{\rm axial}& = & \Tr \intd\tilde c  \epsilon^{\mu \nu \rho\sigma}
G_{\mn} (gA )
G_{\rho\sigma} (gA) + {\cal O} (\epsilon, \epsilonbar) \\
\label{susyanomaly}
\Delta^{\rm susy} 
 & =   & \brs \intd \ln g(x) (L_{\rm YM} + \bar
L_{\rm YM}) \nonumber \\
& = & (\epsilon^ \alpha \delta_\alpha + 
\epsilonbar^ \alphadot \delta_\alphadot) 
\intd \ln g(x) (L_{\rm YM} + \bar
L_{\rm YM}) \nonumber\\
& = & \intd  \Bigl( \ln g(x) i \bigl(\partial _\mu \Lambda ^
\alpha \sigma ^ \mu_{\alpha \alphadot} \epsilonbar^ \alphadot -
\epsilon^ \alpha \sigma^ \mu_{\alpha \alphadot}  \partial_\mu \Lambda
^ {\alphadot} \bigr) \nonumber \\
& & \phantom{\intd} - \frac 12 g^ 2 (x) (\epsilon \chi + \chibar
\epsilonbar)
(L_{\rm YM} + \bar L _{\rm YM}) \Bigr)\ .
\end{eqnarray}
Here
 $L_{\rm
YM}$ and $\Lambda^\alpha$ are the   $F$- and the spinor-component
of the chiral multiplet  $\L_{\rm YM}$ (\ref{LYM}), respectively:  
\begin{equation}
\L_{\rm YM} = - \frac 12 \lambda_a \lambda_a + \theta ^\alpha \Lambda
_\alpha + \theta^2 L_{\rm YM}\ .
\end{equation}

The supersymmetry anomaly is different from the Adler-Bardeen anomaly
 in the following respect:  If gauge invariance is imposed, $\Delta^{\rm
 axial}$ cannot be given as a BRS variation of a renormalizable field
 monomial. In contrast,
 the supersymmetry anomaly is a BRS variation, but it cannot be
 absorbed into the counterterm action, since it depends on the
 logarithm of the coupling. The perturbative loop expansion is a power
 series in the coupling and, thus, the coefficient of the anomaly is
 determined by regularization    scheme- and gauge-independent
 one-loop integrals \cite{KR01anom}.

For Super-Yang-Mills theories the anomaly coefficients $r^{(1)}$ and
 $r_\eta^{(1)}$
 are determined by
\begin{eqnarray}
\label{anomalycoeff}
& &  r^ {(1)} = - \frac {1} {16 \pi^ 2} T(R)\ , \qquad
 r_\eta^{(1)} =  \frac {1} {8 \pi^2} C(G)\ ,
\end{eqnarray}
where $T(R){\mathbf 1}=T_a T_a$ is the Dynkin index of the matter representation
and $C(G) \delta_{ab}= f_{acd} f_{bcd}$ 
is the quadratic Casimir of the adjoint representation.

For the construction of higher orders it is crucial  that
both anomalies can be absorbed into a redefined Slavnov-Taylor
identity,    which defines the 1PI Green functions of the extended model
to all orders of
perturbation theory:
\begin{equation}
\label{STanomalous}
 {\cal  S}^{r_\eta} (\Ga) + r^{(1)} \delta
{\cal S}   \Ga =  0 \ .
\end{equation}
Here $\delta {\cal S}$ is the operator of  the Adler-Bardeen anomaly:  
\begin{eqnarray}
\label{STanomalousoperator}
\delta {\cal S}\Ga 
& \equiv & - 4 i  \intd  \Bigl(\tilde c \Bigl(\frac {\delta}{\delta \eta} -
\frac {\delta}{\delta \etabar}\Bigr) + 2i (\epsilon \sigma^ \mu )^
\alphadot V_\mu \frac {\delta}{\delta \chibar^ \alphadot} 
 - 2i ( \sigma^ \mu \epsilonbar )^ \alpha V_\mu \frac
 {\delta}{\delta \chi^ \alpha} \nonumber \\  & & \phantom{\intd}
  + 2 \epsilonbar_\alphadot \lambdaVbar ^\alphadot 
\frac {\delta}{\delta f} - 2 \lambdaV^\alpha \epsilon_\alpha \frac {\delta}{\delta
 \fbar} \Bigr) \Ga  = - \Delta^{\rm axial} + {\cal O} (\hbar )\ ,
\end{eqnarray}
and in ${\cal S}^ {r_\eta} (\Ga)$ the classical Slavnov-Taylor
operator
(\ref{SToperator}) is supplemented by a  differential
 operator  that expresses the 
supersymmetry anomaly  
\begin{eqnarray}
\label{STreta}
{\cal S}^ {r_\eta} (\Ga) \equiv {\cal S}(\Ga) & - & \intd \biggl( \delta F(g^ 2)(\epsilon
\chi + \chibar \epsilonbar) \Bigl(g^4 \df {g^2} + M_\lambda \df {f} +
M_\lambda \df{\fbar} \Bigr) \nonumber \\ & & \quad - 
i \frac {\delta F }{1+ \delta F}
 \partial_\mu g^ 2 \Bigl(\bigl (\sigma ^ \mu \epsilonbar)^\alpha
\df {\chi^ \alpha}  + (\epsilon \sigma^ \mu )^ \alphadot \df {\chibar^
 \alphadot} \Bigr)\biggr) \ ,\nonumber \\
= {\cal S} (\Ga) & - & r_\eta^ {(1)} \Delta^ {\rm susy} + {\cal O}
{(\hbar^2)}. 
\end{eqnarray}
The function $\delta F$ is a power series in $g^2$. Its lowest order
is renormalization-scheme  independent and unambiguously  determined
by the anomaly 
coefficient
(\ref{anomalycoeff}):
\begin{equation}
\label{deltaFdef}
\delta F(g^2)  = r^{(1)}_\eta g^2  + {\cal O}(g^4) \ .
\end{equation}
   Higher orders are scheme-dependent and depend on the normalization
conditions for the coupling,
 or,
vice versa, once a specific form for $\delta F(g^2)$ is chosen,    the
 normalization of the coupling is determined.
 We want to mention already here
that the choice
\begin{equation}
\delta F = \frac { r_\eta^{(1)}g^2 } {1-  r_\eta^ {(1)}g^2 }\ ,
\end{equation}
yields the NSVZ expression for the gauge $\beta$ function
 \cite{NSVZ83},
 whereas
\begin{equation}
\delta F =  g^2 r_\eta^{(1)} 
\end{equation}
results in a  strictly two-loop $\beta$ function 
in Super-Yang-Mills without matter (cf.~(\ref{betafunctions})).

The operator ${\cal S}^{r_\eta}(\Ga)$ can be considered as a modification
of the supersymmetry transformation of the local gauge coupling
($ F(g^2) = 1 + \delta F(g^2)$):
\begin{eqnarray}
\label{brsanomalous}
 \brs^{r_\eta} g^ 2 &=& - g^ 4 (\epsilon^
\alpha \chi_\alpha + \chibar_\alphadot \epsilonbar^ \alphadot) F(g^
2) - i \omega^\nu\partial_\nu g^2\ , \nonumber \\
 \brs^{r_\eta} (\eta- \etabar)& = & (\epsilon^
\alpha \chi_\alpha - \chibar_\alphadot \epsilonbar^ \alphadot)
- i \omega^\nu\partial_\nu (\eta - \etabar)\ ,
\nonumber\\
\brs^{r_\eta} \chi^ \alpha & = & 2 \epsilon^ \alpha \Big(f + \frac
{M_\lambda} {g^2}\Big)+ i (\sigma^  \mu
\epsilonbar)^ \alpha 
\Bigl(\partial_\mu g^{-2} \frac 1 { F (g^2)} + \partial_\mu (\eta -
\etabar)\Bigr) 
- i \omega^\nu\partial_\nu \chi^\alpha \ , \nonumber \\
\brs^{r_\eta} f &= & M_\lambda (\epsilon^
\alpha \chi_\alpha + \chibar_\alphadot \epsilonbar^ \alphadot) F(g^2)
+  i 
\partial_\mu \chi \sigma^ \mu \epsilonbar - i \omega^ \mu \partial_\mu
f \ . 
\end{eqnarray}
These    modifications are in agreement
 with the supersymmetry algebra, and 
  the anomalous Slavnov-Taylor
 operator (\ref{STanomalous}) and its linearized version have the same
 nilpotency properties as the classical one: 
\begin{eqnarray}
\label{STnilpotency}
& & \bigl( \brs^{r_\eta} _\Ga + r^ {(1)} \delta {\cal S} \bigr)\,
\bigl(\brs^{r_\eta} _\Ga + r^ {(1)} \delta {\cal S}\bigr)
   = 0 \quad \mbox{if} \quad
 {\cal S}^{r_\eta} (\Ga) + r^ {(1)} \delta {\cal S} \Ga = 0 \
,\nonumber \\
& &\bigl (\brs^{r_\eta} _\Ga + r^ {(1)} \delta {\cal S} \bigr)\, 
\bigl({\cal S}^{r_\eta}  + r^ {(1)} \delta {\cal S}\bigr)(\Ga)  = 0 
\quad \mbox{for any functional $\Ga$.} 
\end{eqnarray}

 The anomalous Slavnov-Taylor
identity
 (\ref{STanomalous}) and the identity (\ref{holomorph}) are the
defining symmetry identities for higher-order Green functions. Due to
their nilpotency properties (\ref{STnilpotency}), algebraic renormalization can be
performed to all orders as usually. In contrast,    a construction based on
the usage of invariant regularization schemes will fail in  
higher orders. In particular, it is apparent that a  local
renormalizable action solving the anomalous Slavnov-Taylor identity
 cannot exist. 

From the renormalized Green functions of the extended model one
obtains the Green functions of
softly broken SYM theory by taking  the limit of constant
coupling. Denoting the vertex functional of the theory with constant
coupling by $ \Ga^ {\rm sSYM}$, we have
\begin{equation}
\lim_{G \to g} \Ga \Big|_{\phi^V,\; {\mathbf q},\; {\mathbf\qbar} = 0 } =  \Ga ^ {\rm
sSYM} \ .
\end{equation}

\subsection{Invariant counterterms}
  
\label{sec:invct}

As opposed to the spurion field formalism used in \cite{HKS00},
the present construction
yields all restrictions on the  divergency structure of softly
broken SYM theories.
 Invariant
counterterms of the quantized model
are invariant with respect to  the classical Slavnov-Taylor operator,
the Ward operator ${\cal W}^R$  
and satisfy the identity (\ref{holomorph}):
\begin{equation}
\label{invdef}
\brs_{\Gacl} \Ga_{\rm ct, inv} = 0\ , \quad 
{\cal W}^R\Ga_{\rm ct,inv} = 0\ , \quad
\intd \Bigl(\df {\eta} -
\df{\etabar} \Bigr)
\Ga_{\rm ct, inv} =0  \ . 
\end{equation}
Furthermore, owing to the identification of $\eta+\etabar$ with the  
gauge coupling the formula (\ref{topfor}) holds.
For  the physical, gauge-independent  counterterms, these restrictions
are nothing  but invariance with respect to  the classical symmetries.
They
result therefore in the counterterm action
 (\ref{Gactphysgen}) with the constraints (\ref{zrestricted}) and
 imply the
  non-renormalization theorems and the relations
between the renormalization constants of soft-breaking parameters and
supersymmetry parameters stated already after eq.\ (\ref{zrestricted}).  

However, owing to the non-linearity of the Slavnov-Taylor operator 
the  constraints (\ref{invdef}) 
give rise 
in addition to
field redefinitions for the individual propagating
fields of the theory:
\begin{eqnarray}
\label{fieldred}
A \to z^{(l)}_A g^{2l} A\ , &\qquad & \lambda  \to z^{(l)}_\lambda
g^{2l}\lambda \ , \qquad c \to z^{(l)}_c g^{2l} c \ , \nonumber \\
\varphi_L \to z^{(l)}_\varphi g^{2l} \varphi_L\ ,
& \qquad & \varphi_R \to z^{(l)}_\varphi g^{2l} \varphi_R \ ,\nonumber
\\
\psi_L\to z^{(l)}_\psi g^{2l} \psi_L\ , &\qquad 
&\psi_R \to z^{(l)}_\psi  g^{2l}\psi_R \ .
\end{eqnarray}
These field redefinitions
 are supplemented by external field redefinitions  to
$\brs_\Ga$-invariant expressions
(see \cite{KR01anom} for explicit expressions).

For the softly broken model also 
 the following generalized field redefinitions are relevant:
\begin{eqnarray}
\label{fieldredgen}
& & \lambda_\alpha \to \lambda_\alpha
 + z^{(l)}_{\lambda A} i  g^{2(l+1)}\sigma^\mu_{\alpha
\alphadot}\chibar^ \alphadot A_\mu 
\nonumber 
\\
& & \psi_{X,\alpha} \to \psi_{X,\alpha} + z^{(l)}_{\psi
\varphi}g^{2(l+1)}\chi_\alpha
 \varphi_A\ ,  \quad X = L,R\ .
\end{eqnarray}
The generalized field redefinitions vanish in the limit of constant
coupling,
but when they are
extended to
 $\brs_\Ga$-invariant expressions one obtains two non-vanishing 
 contributions  depending on the gaugino mass:
\begin{eqnarray}
\label{ctextf}
& &  z_{\lambda A}^{(l)}  M_\lambda g^{2l} \intd \big( i
Y_\lambda \sigma^\mu \epsilonbar A_\mu + \mbox{c.c.}\big)\ , \nonumber \\
& & z_{\psi\varphi}^{(l)} g^{2l} M_{\lambda}\intd\big(\epsilon^\alpha
Y_{\psi_L\alpha}
    \varphi_L + \epsilon^\alpha
Y_{\psi_R\alpha}
    \varphi_R + \mbox{c.c.}\big) \ . 
\end{eqnarray}
Their appearance indicates the appearance of corresponding 
additional gauge-dependent divergences in the external field part. However,  
 as they stem from the generalized field
redefinitions,
 their renormalization is irrelevant for
 Green functions of physical fields  (see also \cite{HKS00}).

The invariant counterterms are the only contributions to $\Gamma$  that are not determined
by the Slavnov-Taylor identity (\ref{STanomalous}) and that
  have to be fixed by appropriate normalization conditions.
In order to fix the counterterms to the physical parameters
$z_{g^2}^{(1)}$ and $z_m^{(l)}$,  
a normalization condition for
the coupling in one-loop order and a normalization condition for
the fermion mass have to be used.

The normalization of the coupling is  fixed via the
symmetries for $l>1$, and the normalization of the
soft-breaking parameters is fixed for all $l$ . However, the specific
form of the Slavnov-Taylor identity 
depends on the choice of the function $\delta F(g^2)$ and of the
shifts (\ref{massshifts}). Instead of choosing $\delta F(g^2)$ and the
shifts, it is possible to require independent normalization conditions
also for the coupling in higher orders and for the soft parameters. In  
this case, 
the  coupling
normalization
 determines the terms of order higher than $g^4$ in the function
$\delta F(g^2)$ (\ref{deltaFdef}),
and normalization conditions for soft susy-breaking terms
determine higher-order corrections to the classical shifts:
\begin{eqnarray}
\label{softonshell}
f(x) & \to & f(x) + \frac 1 {g^ 2} \Big({M_\lambda} + \sum _{l=1} ^ \infty v^
{(l)}_\lambda g^ {2l}\Big)\ , \nonumber \\
q_F & \to & q_F(x) - \Big( b   + \sum_{l=1}^ \infty v_b^
{(l)} g^ {2l}\Big)\ , \nonumber  \\
\DV & \to & \DV  - 2 \Big( M + \sum_{l=1}^ \infty v^ {(l)}_M g^ {2l}\Big)^2
\ ,
\end{eqnarray} 

Hence, the present model for the renormalization of soft breakings
gives a consistent description of all renormalization properties of
softly broken supersymmetric Yang-Mills theories. For concrete calculations the
simpler version of spurion fields may be preferred, but the results
on the divergency structure of the softly broken SYM theory    remain valid and
can be used for a consistency check and as a guideline for the
classification of  divergences in concrete diagrams.

\section{The $\beta$ functions of soft-breaking terms}

Using the present construction of softly broken SYM theories, it is
possible to  
determine
the all-order expressions for
the renormalization group $\beta$ functions of the soft mass
parameters.
 As mentioned at the end of sec.\ \ref{sec:RenProperties}, due to
the two anomalies the $\beta$ functions cannot be inferred from the
invariant counterterms but require an algebraic construction.
The derivation we use   is analogous to the one carried out in
softly broken SQED \cite{KRST01soft},
 but it differs in the explicit expressions by the
appearance of the supersymmetry anomaly in the supersymmetry
transformations of the local gauge coupling. 
In the present paper we
skip the construction of the Callan-Symanzik equation. For a more 
detailed discussion of the different partial differential
equations and their relations we refer to \cite{KRST01soft}.

The renormalization group (RG) $\beta$ functions are  uniquely
defined only if one specifies 
the normalization conditions for the parameters of the theory.
 In common usage, the RG $\beta$ functions
are identified with the $\beta$ functions of 
 mass-independent  schemes. As shown in \cite{KR94},   such schemes
correspond to asymptotic 
normalization conditions, where the (Euclidian)
normalization point $\kappa^2$ is
considered as being much larger than the mass parameters of the
theory, and in these schemes
 the  differentiation with respect to the mass parameters is soft
 to all orders.    This property implies
the classical scaling equations for the individual mass differentiations
to all orders:
\begin{eqnarray}
\label{massequ}
 m\partial_m \Ga & = &
 m \intd \Bigl(\frac {\delta} {\delta q} + \frac {\delta} {\delta
\qbar}\Bigr)\Ga   \ ,\nonumber\\
 M_{\lambda} \partial_{M_\lambda}\Ga & = &
  \intd 
\frac {M_\lambda}{g^ 2}\Bigl(\frac {\delta} {\delta f} + \frac {\delta} {\delta
\fbar}\Bigr) \Ga \ , \nonumber \\
M\partial_M \Ga & = & - 4  M^ 2
  \intd \frac {\delta}  {\delta \DV}  \Ga  \ , \nonumber \\
b \partial_b \Ga
 & = &  - b  \intd \Bigl(\frac {\delta} {\delta q_F} + \frac {\delta} {\delta
\qbar_F}\Bigr) \Ga  \  . 
\end{eqnarray}

The RG  equation, which describes the transformation
 of the 1PI Green functions under infinitesimal variations of
the normalization point $\kappa$,    
can in general be expressed as a linear combination of all
invariant operators of the theory. The general basis for linear
operators has been constructed in \cite{KR01} and consists of two
gauge-independent operators ${\cal D}_{\rm kin}$ and
${\cal D}_{Vv}$, and gauge-dependent field
operators 
${\cal N}_{\phi}$. 
  
The gauge-independent operators correspond
  in 
lowest order to the one-loop coupling renormalization and the mass
renormalization given in (\ref{zsusy});
 the gauge dependent operators
represent the field redefinitions of 
eqs.~(\ref{fieldred}) and (\ref{fieldredgen}).
While ${\cal D}_{\rm kin}$, ${\cal D}_{Vv}^{\rm sym}$ are  
invariant operators in the strict sense and commute with the
anomalous
Slavnov-Taylor operator for any functional $\Ga$,
\begin{equation}
\label{Dinv}
\bigl( \brs^{r_\eta}_\Ga + r^{(1)}\delta {\cal S}\bigr) {\cal D} \Ga =
{\cal D}  \bigl({\cal S}^{r_\eta}(\Ga) + r^{(1)}\delta {\cal S}
\Ga\bigr)\ ,
\end{equation}
the gauge-dependent field operators commute only up to
field monomials linear in the propagating fields. 

Using these operators, the RG equation can be written as  
\begin{eqnarray}
\label{RGequ}
& &\biggl( \kappa \partial_\kappa 
+ \hat \beta_{g^2}^{ (1)} {\cal D}_{\rm kin} 
 - \sum_{l\geq 1}\Bigl(\hat \gamma ^ { (l)} {\cal D}^ {{\rm sym}
(l)}_{Vv}  +  \hat \gamma_\varphi^ { (l)} {\cal N}_\varphi^ {(l)} 
+ \hat \gamma_\psi^ {(l)} {\cal N}_\psi^ {(l)} \nonumber \\
& &  \quad
+\hat \gamma_{\psi\varphi}^ { (l)} {\cal N}_{\psi\varphi}^ {(l)}
+\hat \gamma_A^ { (l)} {\cal N}_A^ {(l)}  
+\hat \gamma_\lambda^ { (l)} {\cal N}_\lambda^ {(l)}  
+\hat \gamma_{A\lambda}^ { (l)} {\cal N}_{A\lambda}^ {(l)}  
\Bigr)\biggr) \Ga  =  \Delta_Y \ ,
\end{eqnarray}
where $\Delta_Y$ summarizes the terms linear in the propagating
fields. Its most
important contributions 
are the field monomials of eq.~(\ref{ctextf}),  which belong to the
generalized field 
redefinitions and which are non-vanishing in the limit of constant
coupling.

For the $\beta$ functions of soft mass parameters
only the physical operators $ {\cal D }_{\rm
kin} $ and ${\cal D}^ {\rm sym}_{Vv}$ are relevant.
The explicit form of ${\cal D}_{\rm kin}$ can be obtained from the  
result presented in \cite{KR01} by taking into account the 
gaugino-mass shift in the $F$ component of the
supercoupling. It takes the following form:
\begin{equation}
{\cal D}_{\rm kin} = \intd  F(g^2) \biggl( g^4 \df {g^2} +  
 {M_\lambda}   \Bigl( \df{f} + \df{\fbar} \Bigr) \biggr) \ .
\label{DkinDef}
\end{equation}
The function
$F(g^2) = 1 + \delta F(g^2) = 1 + r_\eta^ {(1)} g^2 + {\cal O}(g^4)$
 is the same function that appears in the anomalous Slavnov-Taylor
 operator (\ref{STanomalous})
and that    absorbs the anomalous supersymmetry-breaking (see (\ref{STreta})). 

The explicit form of the invariant operator $ D^ {{\rm sym} (l)}_{Vv}$
can be found in \cite{KR01}. It has the decomposition  
\begin{equation}
\label{DvVsym}
{\cal D}^ {{\rm sym}(l)}_{Vv} =
{\cal D}^ {(l)}_{Vv} -  8 r^ {(1)}\bigl(  {\cal D}^ {(l+1)}_{g^2} +
 l \big( {\cal N}^ {(l+1)}_V -  8 (l+1)r ^{(1)} \delta {\cal N}^ {(l+2)}_V \big)
 \bigr) .
\end{equation}
Here the operator
 ${\cal D}^ {(l)}_{Vv}$ is invariant with respect to $\brs_\Ga^
{r_\eta}$, i.e,
\begin{equation}
\label{DvVinv}
 \brs^{r_\eta}_\Ga {\cal D}_{Vv} \Ga =
{\cal D}_{Vv} \, {\cal S}^{r_\eta}(\Ga) \ ,
\end{equation}
and the additional terms continue ${\cal D}_{Vv}  $ to an invariant
operator with respect to the complete anomalous Slavnov-Taylor
 operator as defined
in eq.~(\ref{Dinv}).
  
It turns out that the operators composing $D_{Vv}^{\rm sym}$
 depend on the components of the anomalous multiplet $\tilde G$ of the
gauge coupling $g(x)$,
\begin{equation}
\tilde G (x, \theta, \thetabar) = g(x) + {\cal O}(\theta,
\thetabar)\ ,
\end{equation}
which is defined as the vector multiplet with
respect to the anomalous supersymmetry
transformations of the coupling (\ref{brsanomalous}). Its components
are defined by the equation  
\begin{eqnarray}
\label{Greta}
\delta^{r_\eta}_\alpha \tilde G & =& \Bigl(\frac {\partial} {\partial \theta^\alpha} +
i
(\sigma^\mu \thetabar)_\alpha \partial_\mu\Bigr) \tilde G \ , \qquad
\bar \delta^{r_\eta}_\alphadot \tilde G  = \Bigl( \frac \partial
{\partial \thetabar ^\alphadot} - 
i ( \theta \sigma^\mu)_\alphadot \partial_\mu \Bigr) \tilde G\ . 
\end{eqnarray}

 For the
purpose of the present paper we only want to study the limit to
constant coupling in the presence of soft mass shifts.
In  this limit only the $F$ and $D$ 
components of $\tilde G$
are  
non-vanishing, 
\begin{equation}
\lim_{G\to g} \tilde G^ {2l}= g^{2l} \Big(1+ \theta^2 f^{(l)} + \thetabar^2 f^{(l)} +
\frac 14 \theta^2 \thetabar ^2 d^{(l)}\Big)\ ,
\end{equation}
and we obtain as contributions in loop order $l$:
\begin{eqnarray}
f^{(l)} & = & -l M_\lambda 
F(g^2)\ , \nonumber \\
d^{(l)}&= & 4 l(l+1)  M_\lambda^
2
F^ 2 (g^2) + 4 l  M_\lambda^2 g^2 \partial_{g^2} F^2(g^2)  \ .
\end{eqnarray}

For constant coupling and for vanishing external fields $\phi^V$,
 $\mathbf q$, $\mathbf \qbar$, the   
 general expressions of ref.~\cite{KR01} yield
 the following contributions to the individual operators of
 $D_{Vv}^{\rm sym} $ (\ref{DvVsym}):
\begin{eqnarray}
\label{DvVconstant}
{\cal D}_{Vv}^{(l)}  &\, \to \,& g^ {2l} \intd 
\biggl( d^{(l)}\df {\DV}  - 2 
m\Big(\df{q} + \df {\qbar}\Big) 
+ 2 (b  - 2 m f^ {(l)})\Bigl( \df{q_F} + \df{\qbar_F}
\Bigr)\biggr) \ ,
 \nonumber \\
{\cal N}_{V}^{(l )}   &\, \to \,& - 2   M^2 
F(g^2) g^ {2l}\intd 
  \df{\DV}   \ ,  \nonumber \\
{\cal D}_{g^2}^{(l+1)}  &\, \to \,& 
    g^{2(l+2) }   F(g^2) \partial_{g^2} 
- g^{2l}\bigl(f^ {(l)} - M_\lambda   F(g^2)\bigr) \intd  
 \Bigl(\df{f} + \df{\fbar}\Bigr)\ ,   \nonumber \\
\delta {\cal N}_{V}^{(l+2)}  &\, \to\,  & 0\ .
\end{eqnarray}
 For constant coupling it is possible to simplify the operators in
(\ref{DvVconstant}) and ${\cal D}_{\rm kin}$ in (\ref{DkinDef})
further. Using the mass equations (\ref{massequ}) we can eliminate the
field differentiations appearing in the expressions of
(\ref{DvVconstant}) and (\ref{DkinDef}) in favor of mass
differentiations. When the emerging results for the invariant
operators are put into the RG equation (\ref{RGequ}), we obtain the
usual form of a RG
equation for the 1PI Green functions of softly broken
Super-Yang-Mills theories $\Gamma^{\rm sSYM}$:  
%
\begin{eqnarray}
\label{RGconst}
& & \biggl(\kappa \partial_{\kappa} +
 \beta_{g^2} \partial_{g^2} 
+ \beta_{M_\lambda} \partial_{M_\lambda}
+ \beta_{M} \partial_{M}  + \beta_b
\partial_b
+\beta_m \partial_{m}   \nonumber \\
& & 
\quad
 - \sum_{\rm field \ red.} \gamma_{\phi} {\cal N}_\phi
  \biggr)\,   \Ga^ {\rm sSYM}  =  \Delta_{Y} \ .
\end{eqnarray}
The construction determines the $\beta$ functions appearing here
completely: 
The $\beta$ function of the supersymmetric mass parameter $m$
is determined by the gauge-independent coefficients
 $\hat \gamma$ of ${\cal D}_{Vv}$ in (\ref{RGequ}),
\begin{eqnarray}
\label{betam}
\beta_m & = & 2 \gamma \ ,\quad \mbox{with} \quad
 \gamma \equiv \sum_l g^{2l} \hat
\gamma^{(l)}\ ,
\end{eqnarray}
and the $\beta$ function of the gauge coupling is given
as  an all-order expression:
\begin{eqnarray}
\label{betafunctions}
\beta_{g^2}  & = & g^4 \Big(\hat \beta_{g^2}^{(1)}  + 8 r^{(1)} \gamma\Big)
 F(g^2)\ .
\end{eqnarray}
In order to obtain the familiar all-order expressions for the $\beta$ functions
of soft supersymmetry-breaking parameters 
 we rewrite the explicit dependence on the loop order $l$ in the
operators of eq.~(\ref{DvVconstant}) using a
differentiation $\partial_{g^2}$ acting on $\gamma$
(\ref{betam}) and find
\begin{eqnarray}
\label{betafunctionssoft}
\beta_{M_\lambda}  & = &  M_\lambda  F(g^2) g^2 \partial_{g^2}
\Bigl( g^2 \big(\hat \beta_{g^2}^{(1)}  + 8 r^{(1)}  \gamma\big)\Bigr) ,  
\nonumber \\ 
\beta_{M} & = & M \biggl(\frac  { M_\lambda^ 2 F^2(g^2)}
{ M^2} g^2\partial_{g^2} \bigl(g^2 \partial_{g^2} \gamma\bigr) + \frac
{ M_\lambda^ 2 F^2(g^2)}{M^ 2} g^2\partial_ {g^2} \gamma \bigl(1 + g^2
\partial_{g^2} \ln F(g^2) \bigr) \nonumber \\
& & \quad +\, 4 r^{(1)} g^2 F(g^2) g^2 \partial_{g^2}
\gamma
 \biggr) \ ,\nonumber \\
\beta_b & = & 2b \biggl( \gamma  + 2 \frac { M_\lambda F(g^2) m}b g^2
\partial_{g^2} \gamma \biggr) 
\ . 
\end{eqnarray}
In this way,  all $\beta$ functions are determined in a closed
form. 
They depend on the
one-loop $\beta$ function $\hat \beta^{(1)}_{g^2}$, the anomalous dimension of the
supersymmetric mass $\gamma$, the anomaly
coefficient of the Adler-Bardeen anomaly $r^{(1)}$ and the function $F(g^2)$,
which includes in  order $g^2$ the scheme-independent coefficient of
the supersymmetry anomaly $r_\eta^{(1)}$ (\ref{anomalycoeff}).

In higher orders
the soft $\beta$ functions (\ref{betafunctionssoft})
are not equivalent to the ones quoted in the
literature \cite{YA94,HISH98,JAJO97gaugino, AKK98}. 
But it is well-known that $\beta$ functions are
  scheme-dependent\footnote{For 
scheme-dependence of soft $\beta$ functions in supersymmetric theories 
see e.g.~ref.\ \cite{MAVA93}.}.
Indeed, the usual $\beta$ functions are obtained by a finite
redefinition of the gaugino mass parameter.
When the shift in the $F$ component of the coupling 
is modified according to $M_\lambda \to \frac 1 {F(g^2)}M_\lambda$
(cf.~eq.~(\ref{softonshell})),  the gaugino-mass
equation takes  the form
\begin{equation}
\label{masslambdaequ}
M_\lambda \partial_{ M_\lambda} \Ga = 
  \intd 
\frac {M_\lambda}{F(g^2)g^ 2}\Bigl(\frac {\delta} {\delta f} + \frac {\delta} {\delta
\fbar}\Bigr) \Ga = 0 
\end{equation}
instead of the form in (\ref{massequ}),   
and the redefined shift with its soft mass equation defines just another
mass-in\-de\-pen\-dent scheme. 

In the redefined scheme, the $\beta$ functions change according to
$M_\lambda\to M_\lambda/F(g^2)$ and hence take the form  
\begin{eqnarray}
\label{betasoft}
\beta_{M_\lambda}  & = & M_\lambda   g^2 \partial_{g^2}
\Bigl(g^2 F(g^2) \big(\hat \beta_{g^2}^{(1)} + 8 r^{(1)} \gamma\big)\Bigr) = M_\lambda   g^2 \partial_{g^2} \left(
 \frac {\beta_{g^2}} {g^2} \right),  \nonumber \\ 
\beta_{M} & = & M \biggl(\frac  { M_\lambda^ 2 }
{ M^2} g^2\partial_{g^2} \bigl(g^2 \partial_{g^2} \gamma\bigr) + \frac
{ M_\lambda^ 2 }{M^ 2} g^2\partial_ {g^2} \gamma \Bigl(1 + g^2
\partial_{g^2} \ln F(g^2) \Bigr) \nonumber \\
& & \quad +\, 4 r^{(1)} g^2 F(g^2) g^2 \partial_{g^2}
\gamma
 \biggr) \ ,\nonumber \\
\beta_b & = & 2b\bigg( \gamma  + 2 \frac { M_\lambda  m}b g^2
\partial_{g^2} \gamma \bigg)
\ . 
\end{eqnarray}
The $\beta$ functions of the gaugino mass and the $b$ parameter
correspond now exactly to the expressions quoted in the literature 
(see \cite{YA94,HISH98,JAJO97gaugino, AKK98}).
 For the $\beta$ function
of the scalar mass $M$ we find a closed 
 expression in terms of $\gamma$ and $F(g^2)$ 
generalizing the scalar-mass $\beta$ functions of
\cite{KKZ98,JAJO98scalar2}
 to arbitrary
mass-independent normalization conditions 
of the gauge coupling.

The scalar-mass $\beta$ function is particularly interesting.
Naive supergraph arguments fail to produce the correct form of  
$\beta_M$, since a certain contribution called $X$ term 
 \cite{JAJO98scalar} is missing. The $X$ term is defined as the additional 
non-invariant contribution to
the scalar mass $\beta$ function in the parametrization
\begin{equation}
\beta_M = M \frac {M^2_\lambda}{ M^2} \Bigl(g^2 \partial_{ g^2} (g^2 \partial
_{ g^2} \gamma ) + (1 + X )g^2 \partial_{g^2}
\gamma \Bigr)\ ,
\end{equation}
and the present construction determines its value as
\begin{eqnarray}
\label{Xexplicit}
X &= &
g^2
\partial_{g^2} \ln F(g^2)  + 4 \frac {M^ 2}  { M_\lambda^ 2 }
r^{(1)} g^2 F(g^2) 
=  g^2   \Bigl(r_\eta^{(1)}   + 4  r^{(1)} \frac {M^ 2}  { M_\lambda^ 2 }\Bigr)
+ {\cal O}(g ^4) \ .
\end{eqnarray}
with
 $r_\eta^{(1)} = C(G)/8\pi^2$ and $ r^{(1)} = - T(R)/16 \pi^2 $ (see
(\ref{anomalycoeff})).
The result derived here clarifies why the
$X$ term escapes simple superspace arguments:
Since its  lowest-order coefficients  are 
 identified
with the two anomaly 
coefficients of SYM theories, it is evident that the $X$  term itself
cannot be obtained by arguments based on invariant schemes.

The results of the present construction 
 make the notion of the all-order expressions for $\beta$ 
functions more precise by connecting  the all-order
expressions to the conditions on the parameters 
 of the theory. These conditions are encoded in the transformation properties
and mass shifts of external fields which appear in their explicit form
in the anomalous Slavnov--Taylor identity of the extended model
 (\ref{STanomalous}).  
 In particular, 
choosing the function $F(g^2)$
 in the Slavnov-Taylor identity as the
NSVZ function
\begin{equation}
F(g^2) = \frac 1 {1-  r_\eta^{(1)}g^2}\ ,
\end{equation}
and choosing a mass-independent scheme where (\ref{masslambdaequ})
holds, one obtains the well-known NSVZ expressions
 for the gauge-$\beta$ function (\ref{betafunctions}) and for the $X$ term:
\begin{eqnarray}
\beta_{g^2}^{NSVZ} & = & \frac{g^4(\hat\beta^{(1)}_{g^2}+8 r^{(1)}\gamma)}
                              {1-r_\eta^{(1)}g^2}\ ,\\
X^{NSVZ} & =  &
g^2  \frac{  r_\eta^{(1)}   + 4  r^{(1)} \frac {M^ 2}  { M_\lambda^ 2
}}{1-r_\eta^{(1)} g^2}\ .
\end{eqnarray}

\section{Conclusions}
 
The construction of the present paper gives a consistent description
of softly broken Super-Yang-Mills theories including all specific
renormalization properties of supersymmetric field theories. These are
the non-renormalization theorems and the relations between the
renormalization constants of soft-breaking and supersymmetric
parameters.  Both are expressed as relations for the invariant
counterterms 
and in terms of the $\beta$ functions. 
All renormalization properties 
have been derived from
symmetries   of the
classical action of the extended model and their  extension to the 1PI
Green functions. 
Since the classical symmetries are broken by two anomalies -- the
Adler-Bardeen anomaly and a supersymmetric anomaly -- 
 the 1PI Green functions are
characterized by an anomalous Slavnov-Taylor identity. 
It is shown that the two-loop order of the gauge-$\beta$ function as
well as the two-loop contribution of the $X$ term are both induced by 
the anomalies. 

While the present construction embeds SYM theories into an extended
model, where the supersymmetric structure is more appropriately
characterized, it is possible to use the simpler version of the
spurion fields for practical calculations.
  The  results of
this paper remain valid irrespective of the specific scheme and model
one uses for the inclusion of soft breakings. Thus, they
can be used for a systematic classification of divergences and
counterterms in explicit one- and two-loop calculations, and in particular
seemingly accidental  cancellations of divergences can be proven via
the non-renormalization theorems.

In the present paper we have restricted ourselves to a theory
with a simple non-abelian gauge group and one matter
multiplet of charged Dirac fermions in an irreducible representation
of the gauge group. The most important example for such a theory is
supersymmetric QCD.   Generalizations to models with matter multiplets in
reducible representations including chiral trilinear interactions of
supersymmetry with their soft-breaking interactions are
straightforward since no further anomalies will show up. Thus, the
peculiarities of renormalization with local couplings are completely
accounted for in the simple model considered here. However, since the present
construction is based on the different symmetries of the model, a
careful specification of axial symmetries and their axial vector
multiplets has to be done for any  specific model under consideration.

In the present construction we have excluded parity-violating
masses for soft breaking. Introducing them would lead to a non-trivial
change of the extended model. In the spirit of the present
construction they could be introduced 
if we introduce also a photon multiplet with a shifted $D$-component.
Since the photon should be treated as a propagating field, it differs
from  the external axial vectors that are coupled to the parity-even
combination of scalar mass terms. Hence, in the end one has to
 to consider the renormalization of the Fayet-Iliopoulos
$D$-terms \cite{FAIL74}, which 
might induce other properties as the ones derived in the
present paper (cf.~\cite{JJP00,KAVE01}). Closely related 
is the extension to theories with
spontaneous breaking of gauge symmetry like the MSSM. Also in this
case the construction of the present paper has to be generalized.

Finally we are aiming to the renormalization of the complete
MSSM with its non-renormalization theorems. In such a study
the construction of the  
present paper is an important ingredient and  we are convinced that a
construction along its lines  will give important
new insights into the symmetry structure as well as the
renormalization  structure of the MSSM.

\vspace{0.5cm}
{\bf Acknowledgments} 

We thank R. Flume and W. Hollik  for useful discussions.

\newpage

\renewcommand{\theequation}{\thesection.\arabic{equation}}
\setcounter{equation}{0}
\begin{appendix}
\section{The BRS transformations and the Slavnov-Taylor identity} 
 
In this appendix we list the BRS transformations of the propagating
  and
external  fields. Compared to the BRS transformations of \cite{KR01} 
they
include
  the soft mass shifts in the $F$- and $D$-components of external fields.
\begin{itemize}
\item BRS transformations of the gauge vector multiplet and of the
  Faddeev-Popov ghost.
\begin{eqnarray}
\brs A_\mu & = & \frac 1g \partial_\mu g c + i g \big[ A_\mu, c\big]
 + i\epsilon\sigma_\mu\lambdabar
             -i \lambda\sigma_\mu\epsilonbar \nonumber \\
 &  &  +
\frac 12 g ^2(\epsilon \chi + \chibar
\epsilonbar)A_\mu-i\omega^\nu\partial_\nu A_\mu 
\ ,\nonumber \\
\brs\lambda^\alpha & = & - i g \bigl\{ \lambda, c\bigr\} +
\frac{i}{2g} (\epsilon\sigma^{\rho\sigma})^\alpha
             G_{\rho\sigma}(gA) + i\epsilon^\alpha\,
             g(\phibar_L T_a \varphi_L - \varphi_R T_a \phibar_R) \tau_a
 \nonumber \\
& & + \frac 12 \epsilon^ \alpha
g^2 ( \chi \lambda - \chibar \lambdabar) 
+ \frac 12 g ^2(\epsilon \chi + \chibar \epsilonbar) \lambda^ \alpha
              -i\omega^\nu\partial_\nu  
             \lambda^\alpha
\ ,\nonumber \\
\brs c & = & - ig c \, c  + 2i\epsilon\sigma^\nu\epsilonbar A_\nu +
\frac 12 g^2 (\epsilon \chi + \chibar \epsilonbar) c
-i\omega^\nu\partial_\nu c
\ . 
\end{eqnarray}
\item BRS transformations of  matter fields 
\begin{eqnarray}
\brs\varphi_L & = & - i(g c_a T_a - \tilde c)\varphi_L +\sqrt2\,
\epsilon\psi_L
 - i\omega^\nu\partial_\nu \varphi_L
\ , \nonumber \\
\brs\psi_L^\alpha & = & - i(g  c_a T_a - \tilde c)\,\psi_L^\alpha
\nonumber \\
 & &  - \sqrt2\, 
         \epsilon^\alpha\, (\qbar+m)\phibar_R
         -\sqrt2\, i (\epsilonbar\sigmabar^\mu)^\alpha D_\mu\varphi_L 
         -i\omega^\nu\partial_\nu \psi_L^\alpha
\ ,\nonumber  \\
\brs\varphi_R & = &  i \varphi_R (g c_a T_a + \tilde c) +\sqrt2\,
\epsilon\psi_R
 - i\omega^\nu\partial_\nu \varphi_R
\ , \nonumber \\
\brs\psi_R^\alpha & = &  i \psi_R^\alpha (g  c_a T_a + \tilde c)\,
\nonumber \\
 & &  - \sqrt2\, 
         \epsilon^\alpha\, (\qbar+m)\phibar_L
         -\sqrt2\, i (\epsilonbar\sigmabar^\mu)^\alpha D_\mu\varphi_R
         -i\omega^\nu\partial_\nu \psi_R^\alpha
\ .
\end{eqnarray}
\end{itemize}
\begin{itemize}
\item BRS transformations of the axial vector multiplet and of the axial
ghost
\begin{eqnarray}
\brs V_\mu & = & \partial_\mu \tilde c + i\epsilon\sigma_\mu{\bar {\tilde
\lambda}}
             -i \tilde
\lambda\sigma_\mu\epsilonbar -i\omega^\nu\partial_\nu V_\mu
\ ,\nonumber \\
\brs \tilde\lambda^\alpha & = & \frac{i}{2} (\epsilon\sigma^{\rho\sigma})^\alpha
             F_{\rho\sigma}( V) + \frac i2 \epsilon^\alpha\,
             (\tilde D - 2 M^2) -i\omega^\nu\partial_\nu  
             \tilde \lambda^\alpha
\ ,\nonumber \\
\brs \tilde D & = & 2 \epsilon \sigma^ \mu \partial_\mu {\overline {\tilde \lambda}} + 
                  2 \partial_
 \mu  {\tilde \lambda} \sigma^ \mu \epsilonbar 
             -i\omega^\nu\partial_\nu  {\tilde D} \ ,\nonumber \\
\brs\tilde  c & = & 2i\epsilon\sigma^\nu\epsilonbar V_\nu
-i\omega^\nu\partial_\nu\tilde  c
\  .
\end{eqnarray} 
\item BRS transformations of the local coupling and its superpartners
 (\ref{E2def})
\begin{eqnarray}
\brs \eta & = & \epsilon^ \alpha \chi_\alpha -i \omega ^ \nu \partial_\nu \eta \ ,\nonumber \\
\brs \chi_\alpha& = & 2 i (\sigma^ \mu \epsilonbar)_\alpha \partial_\mu \eta + 2
\epsilon_\alpha (f + \frac {M_\lambda} {g^2 }) 
- i \omega^ \mu \partial_\mu
\chi_\alpha\ ,\nonumber \\
\brs f & = & - M_\lambda (\epsilon \chi +\chibar \epsilonbar)+  i 
\partial_\mu \chi \sigma^ \mu \epsilonbar - i \omega^ \mu \partial_\mu
f \ . 
\end{eqnarray}
\item BRS transformations of $q$-multiplets (\ref{qdef})
\begin{eqnarray}
\brs q & = &  + 2i \tilde c (q  + m ) +
 \epsilon^ \alpha q_\alpha -i \omega ^ \nu \partial_\nu q\ ,\nonumber\\
\brs q_\alpha& = & + 2i \tilde c q_\alpha +2 i (\sigma^ \mu
\epsilonbar)_\alpha  
D_\mu q  + 2
\epsilon_\alpha (q_F  -b )
- i \omega^ \mu \partial_\mu
q_\alpha \ ,\nonumber \\
\brs q_F & = &+ 2i \tilde c (q_F -b) +
 i 
D_\mu q^ \alpha \sigma^ \mu_{\alpha \alphadot} \epsilonbar^ \alphadot
- 4 i
 \lambdaVbar _ \alphadot \epsilonbar^ \alphadot (q +m )
 - i \omega^ \mu \partial_\mu
q_F \ . 
\end{eqnarray}
The covariant derivative is defined by
\begin{equation}
\label{covq}
D_\mu { q^ i} = (\partial _\mu - 2 i V_ \mu)({ q^ i} + (m,0,-b))
\end{equation}
\end{itemize}

The classical Slavnov-Taylor identity (\ref{ST})
expresses in functional form BRS invariance of the classical action
and on-shell  nilpotency of BRS transformations.  
 The Slavnov-Taylor operator acting on a general
functional ${\cal F}$  is defined as
\begin{eqnarray}
{\cal S}({\cal F}) & = & 
\intd\biggl\{\dF{\rho_\mu} \dF{A^\mu}
+ \dF{Y_\lambda{}_\alpha}\dF{\lambda^\alpha}
+ \dF{Y_\lambdabar^\alphadot}\dF{\lambdabar_\alphadot}
+   \dF{Y_c} \dF{c}
 \nonumber \\&&{}\quad
         + s B \dF{B} + s\bar{c} \dF{\bar{c}}
\nonumber\\&&{}\quad
+ \Bigl(\dF{Y_{\varphi_L}}\dF{\varphi_L}
+ \dF{Y_{\phibar_L}}\dF{\phibar_L}
+ \dF{Y_{\psi_L{}_\alpha}}\dF{\psi_L^\alpha}
+ \dF{Y_{\psibar_L}^\alphadot}\dF{\psibar_L{}_\alphadot}
+(_{L\to R})\Bigr)
\nonumber\\&&{} \quad
+ s\eta^i\frac{\delta{\cal F}}{\delta\eta^ i}
+ s\etabar^i\frac{\delta{\cal F}}{\delta\etabar^ i}
+ s V^i\frac{\delta{\cal F}}{\delta V^ i}
+ s q^i\frac{\delta{\cal F}}{\delta q^ i}
+ s\qbar^i\frac{\delta{\cal F}}{\delta\qbar^ i}
 \biggr\}
+ s\omega^\nu \frac{\partial{\cal F}}{\partial\omega^\nu} \
. \nonumber \\
\label{SToperator}
\end{eqnarray}
Here we have defined $\eta^i = (\eta, \chi^\alpha, f)$, $V^i =
(V^\mu,\lambdaV, \lambdaVbar,\DV)$ and $q^i = (q, q^\alpha, q_F)$.

\end{appendix}


\end{document}